\documentclass[12pt]{article}
\usepackage{epsf}
\title{ {\bf
The Wigner molecule in a 2D quantum dot}}
\author{\vspace{1cm}\\
         {\bf N. Akman}
         \thanks{E-mail address:
         akman@newton.physics.metu.edu.tr}
         and {\bf M. Tomak}
         \thanks{E-mail address:
         tomak@rorqual.cc.metu.edu.tr}
\\
         Middle East Technical University, Department of Physics, \\
         06531 Ankara, Turkey\\ \vspace{1cm}\\} 

\date{}

\begin{document}
\setlength{\baselineskip}{24pt}
\maketitle
\setlength{\baselineskip}{7mm}
\begin{abstract}
The charge density and pair correlation function of three interacting 
electrons confined
within a two-dimensional disc-like hard wall quantum dot are calculated
by full numerical diagonalization of the Hamiltonian. The formation of a 
Wigner-molecule in the form of equilateral triangular configuration for 
electrons
is observed as the size of the dot is increased.\\


\end{abstract}
\newpage
\section{Introduction}
Advances in nanostructure technology have allowed the lateral confinement of 
a two dimensional electron gas (2DEG) by means of 
suitably shaped gate electrodes or by etching techniques [1]. These 
confined systems having a discrete energy spectra are commonly called
zero dimensional systems or quantum dots [2-9]. The 
motion of electrons in a quantum dot is quantized in all three 
spatial directions. However, if the quantization in the vertical direction
is much stronger than the quantization in the in-plane directions, a quantum
dot could be treated as disc-like 2D system where electrons have 
significant freedom along x- and y- directions.

Experimentally, the number of electrons confined in a quantum dot
could be varied over a considerable range by changing the gate voltage 
applied. Quantum dots containing as few as 2, 3 or 4 electrons have 
already been realized and investigated by optical absorption experiments 
[6,9]. In these
quantum dots or artificial atoms the Coulomb interaction between the 
electrons is very important for understanding their quantum mechanical
properties. Especially correlations among electrons are crucial since 
their effects influence the spectral [10-14] and transport properties 
[15-17] of quantum dots.

So far, various analytical techniques have been devised for handling the 
electron correlations in quantum dot systems. One of these techniques
include the solution of the many particle Schrodinger equation [18]. In 
this three dimensional approach, interacting electrons in non-parabolic
quantum dot systems have been investigated using the formalism of 
Hylleraas [19] where interelectron coordinates are built into the 
wavefunction explicitly. By comparing Hylleraas and Hartree-Fock
type results, in [18], it is shown how the charge redistributions, 
attributable to
correlation interactions, affect ground and excited state energies and 
electron confinement in isolated and coupled two quantum dot 
systems. One- and two- electron ground state energies of a silicon sphere 
embedded in an amorphous silicon dioxide matrix [20] have been calculated 
as a function of sphere size. The electron-electron interaction and 
polarization effects in that study have been treated by perturbation.

A great deal of interest has also gone into analytical investigation of
correlation effects in 2D quantum dot systems. Comparison of energies, 
pair correlation functions, and particle densities of the singlet 
and triplet ground state of quantum dot helium in a magnetic field has been 
investigated in [14] by using Hartree, 
Hartree-Fock, and exact diagonalization methods. The HF results for the 
triplet ground state were found to be in very good agreement with the exact 
diagonalization results, proving the importance of the exchange interaction.
On the other hand, the exact results for the singlet state have 
disagreements 
with the HF results. In [14] it has been shown that the disagreement 
between two results arises due to the electron correlations
which are neglected in the Hartree-Fock approximation. However, 
most of the work performed on the electron-electron correlations in 
quantum dots placed in a magnetic field [6,8,21], as well as transport 
experiments [22] and far-infrared spectroscopy [23] has been based on a two 
dimensional quantum dot with parabolic type confinement potential.

Typically the lateral confinement potential in quantum dots is created
by spatially extended charge distributions. Therefore, it shows a parabolic
characteristic ($\sim r^{2}$) and seems to represent fairly well the 
electrostatically confined electrons. The advantage of using a harmonic type
of confining potential is the analytical simplicity of the problem since 
the center of mass and relative coordinates decouple. However, 
in the far-infrared spectroscopy, the 
radiation field can not detect the electron-electron correlations when 
the confinement is parabolic. This is caused by the 
decoupling of far-infrared radiation with the relative motion of 
electrons [24].

Thus, in order to probe the interaction effects, 
it has been suggested that the shape of the dot
should be modified to achieve the coupling of center of mass and relative 
motions [24]. In [24], the heat capacity results have also been presented
to suggest that the interaction effects could be seen by measuring the
thermodynamic properties of the electrons. The 
magnetization of parabolic quantum dots has been also computed
and magnetization found to be another probe 
of the interaction effects [25] like the heat capacity. The 
magnetization of the dots is predicted to oscillate with magnetic 
field because the ground state prefers to be at certain 
magic values of the total angular momentum. This 
behaviour is explained in a subsequent work
[26] as a direct consequence of the Pauli principle which enables the 
electrons to reduce their energy optimally only at the magic angular 
momenta. Recently, vertically coupled quantum dots or artificial 
molecules [27,28] have attracted considerable attention. In [28], 
a double dot system with three spin-polarized
electrons was investigated and a sequence of angular-momentum magic numbers
was found depending on the strength of the interdot tunneling. Besides 
these, some authors have achieved the coupling of center of mass
and relative motions of electrons with deviations from the exact 
harmonic confinement [29-31]. With etching techniques or 
self organized growth it is possible to create 
hard-wall type confinement potential [32]. Since 
the center of mass motion of electrons is coupled with their
relative part in this type of confinement potential, the effects of 
correlation influencing energy spectrum, transport, and spectral measurements
can be studied with the excitation spectra [33].

In this work we analyze electron correlation effects in a 2D circular
quantum dot with hard wall confinement potential. We 
employ exact diagonalization technique in computing the eigenvalues 
and corresponding eigenvectors for the ground state numerically. We 
specialize to the case of three spin-polarized electrons with 
total magnetic quantum number $S_{z}=3/2$ and total 
orbital angular momentum $M=0$. We investigate charge density and pair 
correlation function as a function of the dot size under these 
circumstances. With increasing dot size the observed 
formation of Wigner molecule in equilateral triangular configuration
of three electrons is discussed. In section 2 we 
give the formalism and details of the calculation. In 
section 3 and 4 we report and discuss the results 
of the numerical calculations.
\section{Model and the method of calculation}
The total orbital angular momentum $M=\sum_{i=1}^{N}m_{i}$,
the total spin S and the total magnetic quantum number 
$S_{z}=\sum_{i=1}^{N}s_{i}$ are good quantum numbers for N electrons in 
the dot due to the circular geometry of the dot, and spin independence of 
the Coulombic interaction. In second quantization language,
the N electron quantum dot is described by the Hamiltonian
\begin{eqnarray}
{\cal{H}}=\sum_{K}{\cal{E}}_{K}a_{K}^{\dagger}a_{K}+\frac{\lambda}{2}
          \sum_{K,L,M,N}
          V_{K,L,M,N}\;a_{K}^{\dagger}a_{M}^{\dagger}a_{L}a_{N}\,\,,
\end{eqnarray}
where ${\cal{E}}_{K}=\frac{1}{2}k_{n_{K},|m_{K}|}^{2}$ depends
only on the $n_{K}$- th root of the Bessel function of the first kind
$J_{m_{K}}$. In writing eqn. (1), length is measured in units of the dot 
radius
a, and energy in units of $\hbar^{2}/(m^{\star}a_{B}^{2}\lambda^{2})$,
where $m^{\star}$ is the electron effective mass determined by the host 
semiconductor. In (1), 
the Coulomb potential is multiplied by $\lambda=a/a_{B}$ which
serves as the dimensionless coupling constant characterizing the 
strength of the interaction. Although it is 
possible to perform a perturbative expansion in powers of $\lambda$
to calculate the energy spectrum [34], we prefer to diagonalize the 
Hamiltonian matrix exactly [33,35]. The Coulomb matrix element is
defined via the one-electron orbitals as 
\begin{eqnarray}
V_{K,L,M,N}= \int \int d^{2}{\bf \vec{x}}  d^{2}{\bf \vec{x}} ^{\prime}
             \varphi_{K}^{\star}({\bf \vec{x}} )
             \varphi_{L}({\bf \vec{x}} )V({\bf \vec{x}} -{\bf \vec{x}}
             ^{\prime})
             \varphi_{M}^{\star}({\bf \vec{x}} ^{\prime})
             \varphi_{N}({\bf \vec{x}} ^{\prime}),
\end{eqnarray}
where $\varphi_{A}(\vec{x})$ (A=K, L, M, N) are the eigenfunctions of the 
single free particle Hamiltonian, and A is a collective index designating 
the radial ($n_{A}$), angular momentum ($m_{A}$) and spin quantum numbers 
($\sigma_{A}$) of the electron.
$\varphi_{A}(\vec{x})$ can be written as 
\begin{eqnarray}
\varphi_{A}({\bf \vec{x}} )=\phi_{n_{A},m_{A}}({\bf \vec{x}} )
\chi_{\sigma_{A}}\,\,,
\end{eqnarray}
where $\chi_{\sigma}$ is the spin wavefunction, and one electron orbital 
$\phi_{n,m}({\bf \vec{x}} )$ has the form
\begin{eqnarray}
\phi_{n,m}({\bf \vec{x}} )=\frac{1}{\sqrt{\pi}}\frac{1}{|J_{|m|+1}
                            (k_{n,|m|})|}
                     e^{im\theta}J_{|m|}(k_{n,|m|}|{\bf \vec{x}} |)\,\,,
\end{eqnarray}
and the corresponding eigenvalue ${\cal{E}}_{K}$ is independent of both 
spin and and sign of $m$. As the one electron orbitals depend already on 
the Bessel functions, it is convenient to expand the Coulomb potential, 
$V({\bf \vec{x}} -{\bf \vec{x}} ^{\prime})=\frac{1}{|{\bf \vec{x}} -{\bf
\vec{x}} ^{\prime}|}$, in terms of the Bessel functions, 
\begin{eqnarray}
\frac{1}{|{\bf \vec{x}} -{\bf \vec{x}} ^{\prime}|}=\sum_{m=-\infty}^{\infty}
             \int_{0}^{\infty}dk\,e^{im(\theta-\theta^{\prime})}
             J_{|m|}(k\rho)J_{|m|}(k\rho^{\prime})e^{-k(z_{>}-z_{<})},
\end{eqnarray}
where $z_{>}-z_{<}$ shows the extension of the dot in the longitudinal 
direction. If the longitudinal extension of the dot is a non-negligible 
fraction of its in-plane size, one is to analyze the excitations in this 
direction, too. However, in the limit of $z_{>}-z_{<}\rightarrow 0$, it 
is sufficient to probe only the transversal plane assuming that the 
system is in the lowest state for longitudinal dynamics. In the 
following, we follow the latter one and ascribe a small value to 
$z_{>}-z_{<}$ in the calculations. In this approximation the 
matrix element of the Coulomb potential becomes
\begin{eqnarray}
&&V_{m_{K},m_{L},m_{M},m_{N}}^{n_{K},n_{L},n_{M},n_{N}}=
4\frac{1}{|J_{|m_{K}|+1}(k_{n_{K},|m_{K}|})|}
\frac{1}{|J_{|m_{L}|+1}(k_{n_{L},|m_{L}|})|}
\frac{1}{|J_{|m_{M}|+1}(k_{n_{M},|m_{M}|})|}\nonumber\\
&&\frac{1}{|J_{|m_{N}|+1}(k_{n_{N},|m_{N}|})|}
\int_{0}^{\infty}dk\,\int_{0}^{1} d\rho\, \rho
J_{|m_{K}|}(k_{n_{K},|m_{K}|}\rho)
J_{|m_{L}|}(k_{n_{L},|m_{L}|}\rho)\nonumber\\
&&J_{|m_{K}-m_{L}|}(k\rho) \int_{0}^{1} d\rho^{\prime}\, \rho^{\prime}
J_{|m_{M}|}(k_{n_{M},|m_{M}|}\rho^{\prime})
J_{|m_{N}|}(k_{n_{N},|m_{N}|}\rho^{\prime})
J_{|m_{N}-m_{M}|}(k\rho^{\prime})\,\,.
\end{eqnarray}
One notes that, angular integrations in $V_{K,L,M,N}$ require 
$m_{K}-m_{L}=m_{N}-m_{M}$ otherwise $V_{K,L,M,N}$ vanishes. 

For calculating the physical quantities such as energy spectrum, charge 
density and pair correlation function, one needs to compute the matrix 
elements of the Hamiltonian operator between N-electron ground state
which can be expressed as 
\begin{eqnarray}
|\Phi_{0}^{(N)}>=\sum_{C} b^{0}_{C}|C>,
\end{eqnarray}
where 
$|C>=a_{n_{1},m_{1},\sigma_{1}}^{\dagger}....a_{n_{N},m_{N},\sigma_{N}}
^{\dagger}|0>$ is a non-interacting Slater determinant
and sum runs over 
all possible configurations of the quantum numbers ($n, m, \sigma$) 
satisfying given values of $M$ and $S_{z}$. The expansion coefficients 
$b^{0}_{C}$ are identified with the eigenvectors of the ground state. 
In this work we are concerned with the quartet ground state 
$|S_{z}=3/2, M=0>$
\begin{eqnarray}
|S_{z}=3/2, 
M=0>=a_{n_{1},m_{1},\uparrow}^{\dagger}a_{n_{2},m_{2},\uparrow}^{\dagger}
a_{n_{3},m_{3},\uparrow}^{\dagger}|0>,
\end{eqnarray}
in which all electrons are spin polarized. We will compute all relevant
quantities in the quartet ground state.

The charge density of the electrons are defined by 
\begin{eqnarray}
\rho(\vec{x})=\sum_{\sigma}<\Phi_{0}^{(N)}|\phi_{\sigma}^{\dagger}(\vec{x}) 
\phi_{\sigma}(\vec{x})|\Phi_{0}^{(N)}>,
\end{eqnarray}
which measures the electron density at a given point $\vec{x}$ in space. 
A close 
inspection of this formula reveals that, at a fixed radius, $\rho(\vec{x}) 
\sim 
\cos(2 M \theta)$ so that angular dependence of the charge density is 
determined by the total orbital angular momentum. Hence, 
for $M=0$ angular dependence disappears and $\rho(\vec{x})$ 
assumes only a radial variation dictated by the associated Bessel 
functions. 

Another relevant quantity, the pair correlation function, is a two-point 
function defined by 
\begin{eqnarray}
\rho_{c}(\vec{x},\vec{x}^{\prime})=
\sum_{\sigma,\sigma^{\prime}}<\Phi_{0}^{(N)}|\phi_{\sigma}^{\dagger}(\vec{x})
\phi_{\sigma^{\prime}}^{\dagger}(\vec{x}^{\prime})
\phi_{\sigma^{\prime}}(\vec{x}^{\prime})
\phi_{\sigma}(\vec{x})|\Phi_{0}^{(N)}>,
\end{eqnarray}
which is the probability of finding an electron at $\vec{x}$ given that 
another one is situated at $\vec{x}^{\prime}$.

In the next section we perform a numerical computation of the charge 
density and pair correlation function to identify their dependence on the 
space coordinates as well as the dot size $\lambda$. 
\section{Numerical Analysis}
Electron-electron correlations have been computed for a 1D hard-wall type 
dot in [36], and a 2D parabolic dot in [37]. In the latter 
the Schroedinger equation is solved for three electrons numerically for 
a fixed dot size. For three electrons with $S_{z}=3/2$ and $M=3k$ 
(k=0,1,2, ...) it is expected that, in the ground state, electrons form an equilateral triangle 
at some radius $r_{0}$ determined by the confining parabolic potential. Here 
we analyze charge density and pair correlation function for three electrons 
in a circular 
quantum dot for $S_{z}=3/2$, $M=0$. In contrast to the parabolic 
confinement potential, in the case of hard-wall type confinement one 
cannot find an analitic expression for $r_{0}$. Despite of this, however, 
for large enough $\lambda$ 
value, the equilateral triangular configuration is expected to occur at some
distance in the radial direction. This then will be an indication of 
the Wigner molecule structure which is a general feature of quantum dots 
for large $\lambda$ [36].

Depicted in Fig.1 is the $r$ and $\lambda$ dependence of the charge 
density $\rho(\vec{x})$ (normalized to N=3 for each $\lambda$)
for the state $|S_{z}=3/2, M=0>$. The three curves in this figure describe 
variations of the electron distribution as a function of $\lambda$ 
and $r$. For small $\lambda$ (e. g. $\lambda=1$) electrons have 
non-negligible
distribution at the center of the dot though maximum is reached away from the
origin. This behaviour of $\rho(\vec{x})$ shows that for small $\lambda$
$J_{m=0}(r)$ is the dominant one especially at small $r$. As $\lambda$ 
increases, 
however, the charge density vanishes gradually around the center of the 
dot (e. g. $\lambda=10, 100$). It is here that one observes the dominance 
of $J_{m>0}$ since they vanish at the origin by definition. Besides
the behaviour of the charge density close to the center of the dot,
one observes that higher the $\lambda$ sharper and farther from the 
center the peaks become. Hence, as $\lambda$ increases electrons get shifted 
towards
the periphery (never reaching there due to the infinite potential barrier) 
with a sharper
peak showing the most probable radial position of the electrons. For 
example, for $\lambda=100$ electrons are most probably distributed along 
the periphery at the radial distance $r\sim 0.7$. That electrons move towards
the periphery with increasing $\lambda$ follows from the fact that they
try to minimize their electrostatic energy which is known to be the 
dominant component for large  $\lambda$. Behaviour of the charge 
density gives information only on the radial distribution of electrons;
therefore, one sould also investigate the behaviour of the  pair correlation 
function to obtain the distribution of the electrons 
in the plane of the dot.

In Fig.2 the variation of the pair correlation function (normalized to 
N(N-1)=6) with $\theta$ and $\lambda$ for $r=r'=0.67$ and 
$\theta^{\prime}=0$ is presented. The pair correlation function
$\rho_{c}(\vec{x},\vec{x}^{\prime})$ gives the probability distribution of
$N-1$ electrons given that one of the $N$ electrons is located at 
$\vec{x}'$. For the case of three electrons, the pair correlation function
in Fig.2 shows the angular distribution of two electrons along the periphery
of the quantum dot. Therefore, the fact that pair correlation function
has always two distinct peaks and vanishes at $\theta=0$ is a restatement
of the properties of the pair correlation function consistent with the Pauli
exclusion principle.

An important property of the pair correlation function follows from its
variation with the dot size in units of Bohr radius. One notices that, for 
small $\lambda$ values peaks are not sharp and electrons do not have a 
well-defined configuration, that is, the pair correlation function is not 
diminished significantly between the two peaks. This small $\lambda$
limit shows nothing but the atomic regime of confinement, where dot size 
is of the order of Bohr radius or smaller, and their average kinetic energy 
exceeds the Coulombic repulsion. It is with the dominant kinetic energy of 
the electrons that they are distributed in the dot without a well-defined
configuration. It is known that in this limit  perturbation theory is a 
reliable tool to investigate the physical parameters of the dot [38],
in other words, Hartree-Fock method can be applied just as in the 
few-electron atoms.

Furthermore, as Fig. 2 shows clearly, when $\lambda$ is increased peaks get
sharper. Pair correlation function gradually assumes two distinct peaks 
between which there is a strong depletion. Therefore, higher the dot size 
smaller the overlap between the two peaks. Indeed, as $\lambda$ increases 
the peaks approach to fixed positions such that their angular seperation 
is $\sim 120 ^{\circ}$. This angular seperation constitutes an equilateral 
triangular 
structure as already emphasized in other investigations too [37,38]. This 
geometrical arrangement of the electrons correspond to the minimal energy 
configuration where their kinetic energy is smaller than the 
electrostatic Coulomb energy. The latter is minimized by an equilateral 
triangular configuration of the electrons. Approach of the configuration to
an equilateral triangular structure for large dot sizes is nothing but
the well-known Wigner molecule structure which has been shown to exist 
in other types of dots too [36,38]. Therefore, for large dot sizes one 
observes
that the electrons form a Wigner molecule, that is, they assume fixed 
positions in the dot minimizing the dominant electrostatic energy.

For a better understanding of the behaviour of the electron distribution 
in the dot it may be convenient to analyze the pair correlation function 
in the $r-\theta$ plane in which both radial and angular variations 
become visible. Fig. 3 shows the contour lines of 
$\rho_{c}(\vec{x},\vec{x}^{\prime})$ in $r-\theta$ plane for 
$\theta^{\prime}=0$, $r'=0.67$, and $\lambda=1$. In accordance with Fig. 1
and Fig. 2 there is a non-negligible correlation between the electrons
for small $r$, also the maxima are reached away from the origin, 
$r\sim 0.4$. Moreover near the periphery of the dot, the correlation 
distribution is highly suppressed. One also notices that the pair 
correlation function in between the two peaks is not suppressed at all.

To illustrate the effects of larger dot sizes on Fig. 3 we show in Fig. 4
the pair correlation function in $r-\theta$ plane for $\lambda=100$. As is 
seen there are important differences between Fig. 3 and Fig. 4. First of 
all, the peaks are pushed towards the periphery of the dot in comparison 
with Fig. 1. Next the peaks are now sharpened and the pair correlation
between them are reduced significantly. In this sense the difference 
between Figs. 3 and 4 shows the crytallization process of the electrons 
with increasing Coulombic repulsion among them with growing $\lambda$. 
\section{Conclusion}
In this work we have performed a detailed numerical study of the electron
correlation effects in a 2D circular quantum dot with hard-wall 
confinement potential. Our investigations show that as the dot size 
increase gradually configuration of the electrons approach an equilateral 
triangular structure. Hard-wall quantum dots, which may be as interesting as
parabolic ones for experimental studies, allow for the formation of the 
Wigner molecule structure of the electrons for large dot sizes 
compared to the Bohr radius. It is known that the optical properties like
the inelastic light scattering or far infrared 
absorption [39] and magnetic properties are all dependent on the spin of 
the ground state. Therefore, experimental studies on the quantum 
dots can reveal valuable information on the electron configuration in the
dot by concentrating on the spin dependent quantities.

\newpage
\begin{center}
{\bf Figure Captions}
\end{center}
Fig. 1. Variation of the charge density $\rho(\vec{x})$ with radial
distance $r$ and the dot size $\lambda$.\\ 
Fig. 2. Variation of the pair correlation function with $\theta$
and $\lambda$ for $r=r'=0.67$ and $\theta'=0$.\\
Fig. 3. Contour lines showing the pair correlation function on 
($r-\theta$) plane for $r'=0.67$, $\theta'=0$, and $\lambda=1$.\\ 
Fig. 4. The same as in Fig. 3 but for $\lambda=100$. 
\newpage
\begin{figure}
\vspace{15 cm}
\end{figure}  
\begin{figure}
\vspace{5.0cm}
    \includegraphics{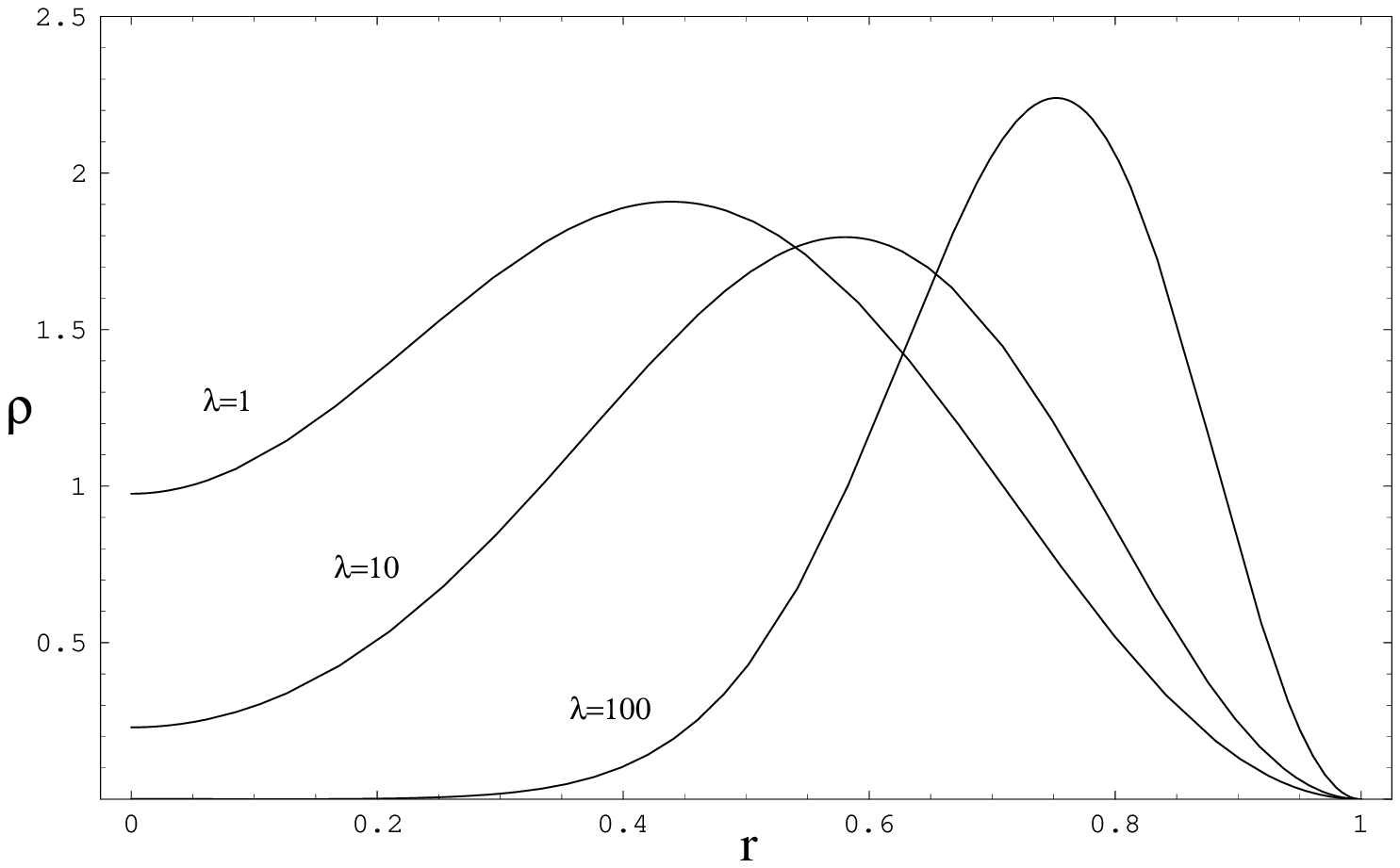}
\mbox{\bf{Fig.1}}
\end{figure}  
\begin{figure}
\vspace{7.0cm}
\end{figure}  
\begin{figure}
\vspace{12.0cm}
    \includegraphics{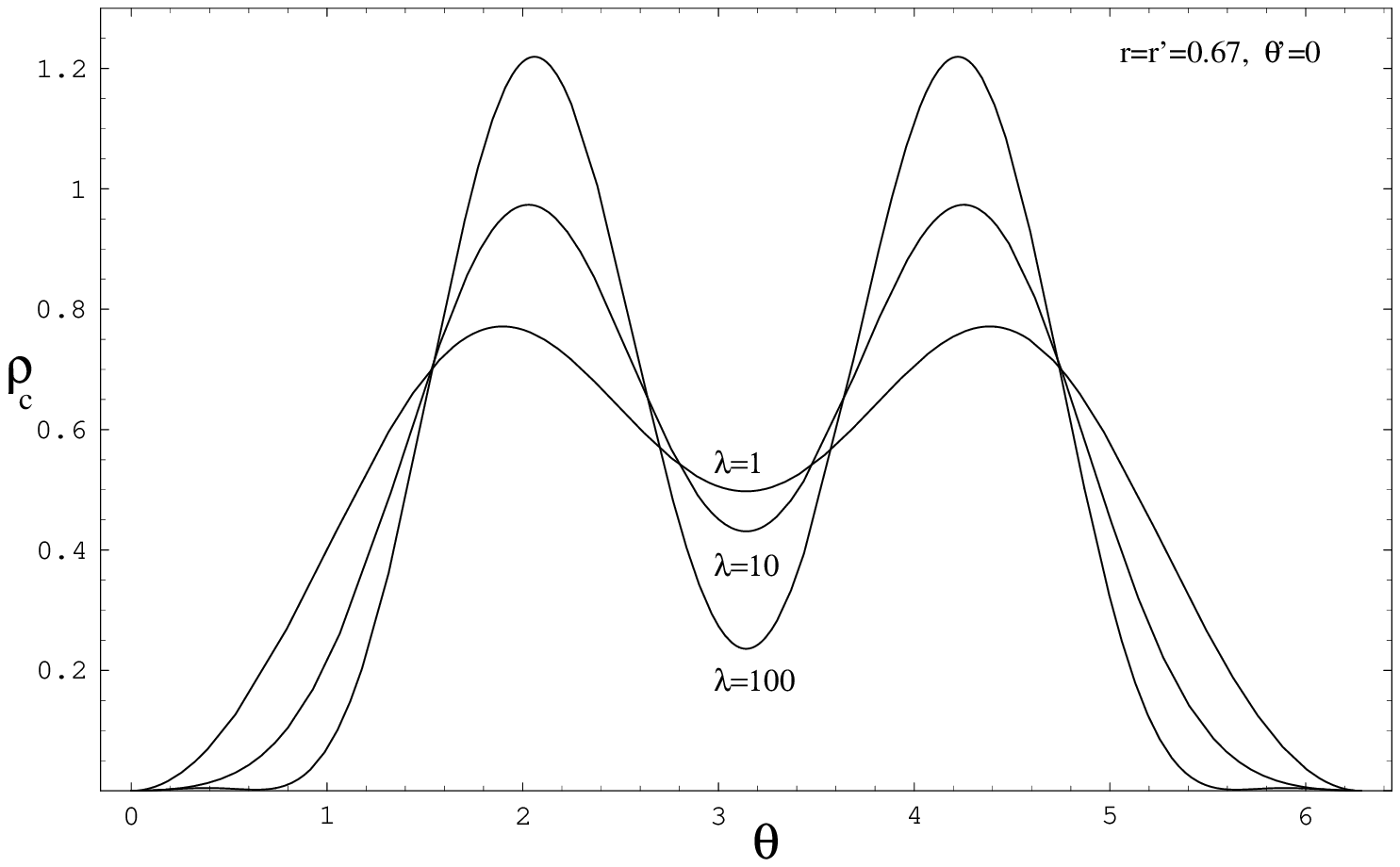}
\mbox{\bf{Fig.2}}  
\end{figure}
\begin{figure}
\vspace{8.0cm}
\end{figure}  
\begin{figure}
\vspace{12.0cm}
    \includegraphics{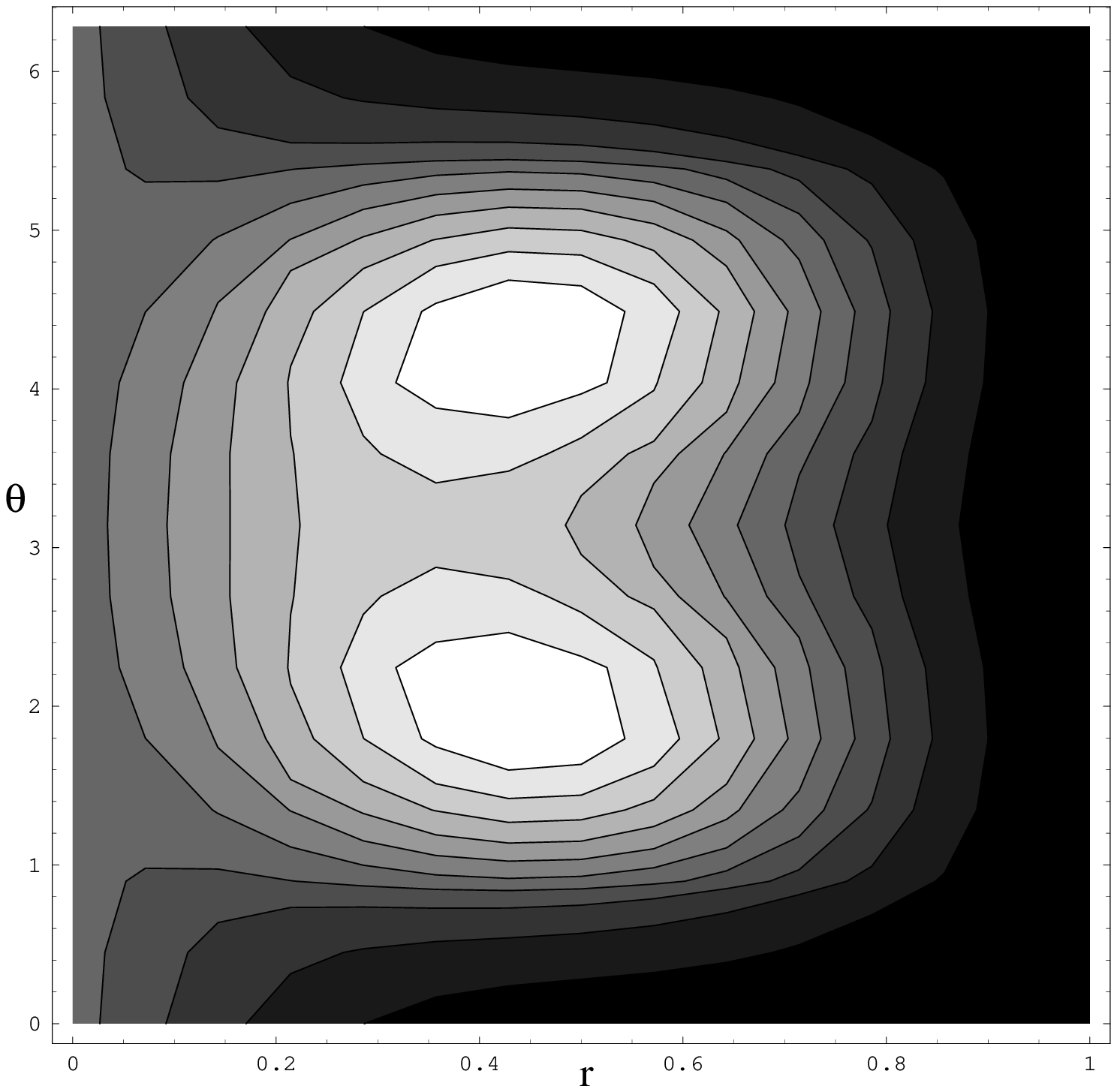}
\mbox{\bf{Fig.3}}
\end{figure}
\begin{figure}
\vspace{8.0cm}
\end{figure}
\begin{figure}
\vspace{12.0cm}
    \includegraphics{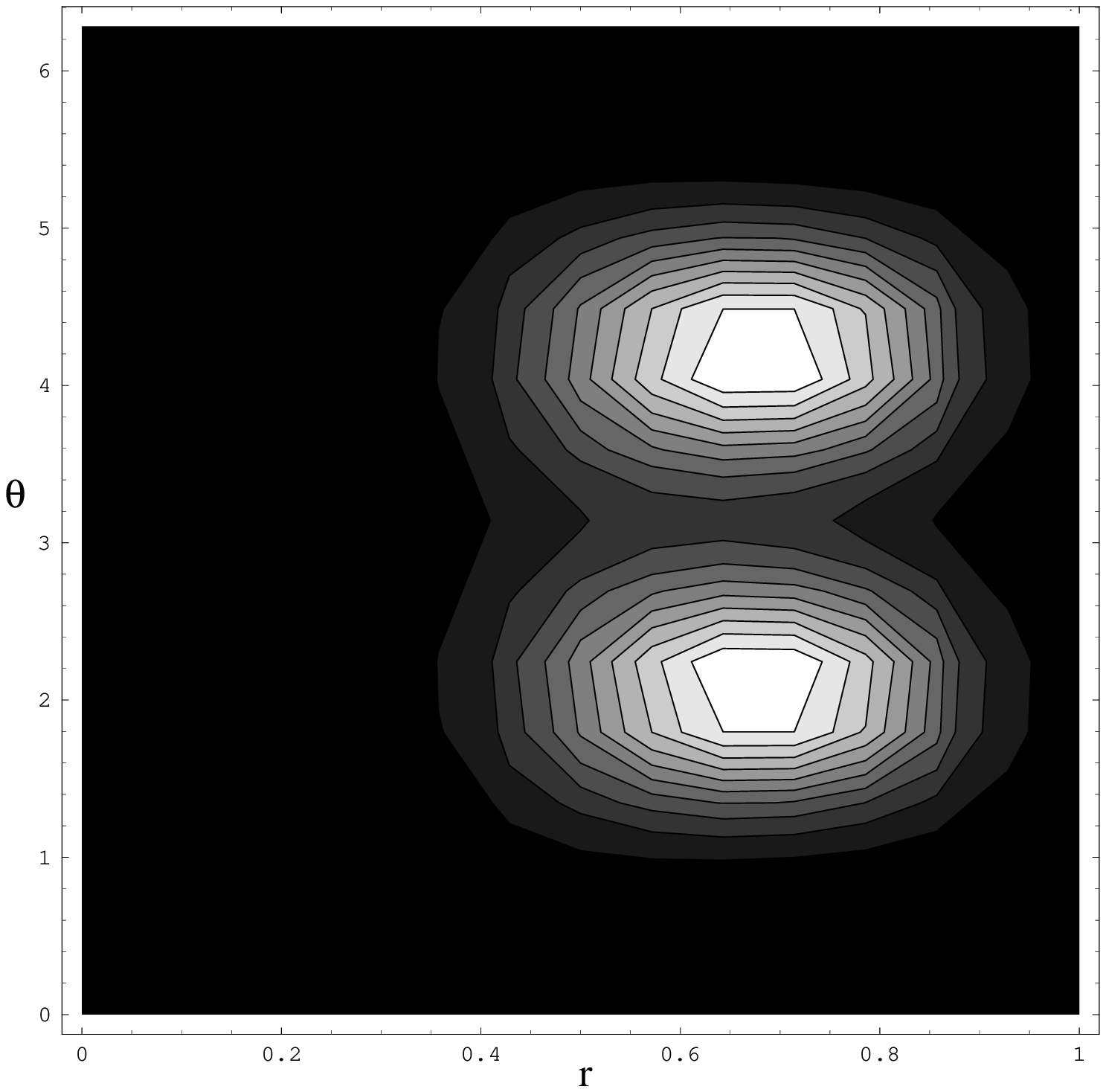} \mbox{\bf{Fig.4}}
\end{figure}
\begin{figure}
\vspace{8.0cm}
\end{figure}

\begin{thebibliography}{99}
\bibitem{1} {G. W. Bryant, Phys. Rev. Lett. {\bf 59} (1987) 1140.} 
\bibitem{2} {M. A. Reed, J. N. Randall, J. Aggarwal, R. J. Matyi, T. M. 
Moore, and A. E. Wetsel, Phys. Rev. Lett. {\bf 60} (1988) 535.}
\bibitem{3} {W. Hansen, T. P. Smith III, K. Y. Lee, J. A. Brum, C. M. 
Knoedler, J. M. Hong, and D. P. Kern, Phys. Rev. Lett. {\bf 62} (1989) 
2168; {\bf 64} (1990) 1991; W. Hansen, T. P. Smith III, K. Y. Lee, J. M. 
Hong, and C. M. Knoedler, Appl. Phys. Lett. {\bf 56} (1990) 168.}
\bibitem{4} {L. P. Kouwenhoven, F. W. J. Hekking, B. J. van Wees, C. J. 
P. M. Harmans, C. E. Timmering, and C. T. Foxon, Phys. Rev. Lett. {\bf 
65} (1990) 361.}
\bibitem{5} {P. L. McEuen, E. B. Foxman, U. Meirav, M. A. Kastner, Y. 
Meir, N. S. Wingreen, and S. J. Wind, Phys. Rev. Lett. {\bf 66} (1991) 
1926.} 
\bibitem{6} {Ch. Sikorski and U. Merkt, Phys. Rev. Lett. {\bf 62} (1989)
2164; {\bf 64} (1990) 3100; Surf. Sci. {\bf 229} (1990) 282.}
\bibitem{7} {A. Lorke, J. P. Kotthaus, and K. Ploog, Phys. Rev. Lett. 
{\bf 64} (1990) 2559.}
\bibitem{8} {T. Demel, D. Heitmann, P. Grambow, and K. Ploog, Phys. Rev. 
Lett. {\bf 64} (1990) 788.}
\bibitem{9} {B. Meurer, D. Heitmann, and K. Ploog, Phys. Rev. Lett. {\bf 68}
(1992) 1371.}
\bibitem{10} {W. Hausler and B. Kramer, Phys. Rev. {\bf B47} (1993) 16353.}
\bibitem{11} {P. Hawrylak and D. Pfannkuche, Phys. Rev. Lett. {\bf 70} (1993)
485.}
\bibitem{12} {U. Merkt, J. Hausler, and M. Wagner, Phys. Rev. {\bf B43} 
(1991) 7320.}
\bibitem{13} {W. Hausler, B. Kramer, and J. Masek, Z. Phys. {\bf B85} (1991) 
435.}
\bibitem{14} {D. Pfannkuche, V. Gudmudsson, and P. A. Maksym, Phys. Rev. {\bf
B47} (1993) 2244.}
\bibitem{15} {Y. Meir, N. S. Wingreen, and P. A. Lee, Phys. Rev. Lett {\bf 
66} (1991) 3048; J. M. Kinaret, Y. Meir, N. S. Wingreen, P. A. Lee,
and X. G. Wen, Phys. Rev. {\bf B46} (1992) 4681.}
\bibitem{16} {J. J. Palacios, L. Martin-Moreno, and C. Tejedor, Europhys. 
Lett. {\bf 23} (1993) 495.}
\bibitem{17} {S. R. Patel, D. R. Stewart, C. M. Marcus, M. Gokcedag, Y. 
Alhassid, A. D. Stone, C. I. Duruoz, and J. S. Harris, Jr, Phys. Rev. Lett
{\bf 81} (1998) 5900.}
\bibitem{18} {A. H. Guerrero, Semi. Sci. Tech. {\bf 10} (1995) 759.}
\bibitem{19} {E. A. Hylleraas, Z. Phys. {\bf 54} (1929) 347.}
\bibitem{20} {D. Babic, R. Tsu, and R. F. Greene, Phys. Rev. {\bf B45}
(1992) 14150.}
\bibitem{21} {P. Bakshi, D. A. Broido, and K. Kempa, Phys. Rev. {\bf B42} 
(1990) 7416.}
\bibitem{22} {P. L. McEuen, N. S. Wingreen, E. B. Foxman, J. Kinaret, U. 
Meirav, M. A. Kastner, and Y. Meir, Physica {\bf B189} (1993) 70.}
\bibitem{23} {U. Merkt, Physica {\bf B189} (1993) 165.}
\bibitem{24} {P. A. Maksym and T. Chakraborty, Phys. Rev. Lett. {\bf 65}
(1990) 108.}
\bibitem{25} {P. A. Maksym and T. Chakraborty, Phys. Rev. {\bf B45} (1992)
1947.}
\bibitem{26} {P. A. Maksym, Physica {\bf B184} (1993) 385.}
\bibitem{27} {B. Partoens, A. Matulis, and F. M. Peeters, Phys. Rev. {\bf 
B59} (1999) 1617.}
\bibitem{28} {H. Imamura, P. A. Maksym, and H. Aoki, Phys. Rev. {\bf 53}
(1996) 12613.}
\bibitem{29} {D. Pfannkuche and R. R. Gerhardts, Phys. Rev. {\bf B44} (1991)
13132.}
\bibitem{30} {V. Gudmundsson and R. R. Gerhardts, Phys. Rev. {\bf B43} 
(1991) 12098.} 
\bibitem{31} {Z. L. Ye and E. Zaremba, Phys. Rev. {\bf B50} (1994) 17217.}
\bibitem{32} {F. M. Peeters and V. A. Schweigert, Phys. Rev. {\bf B53}
(1996) 1468.}
\bibitem{33} {A. Brataas, U. Hanke, and K. A. Chao, Semi. Sci. Tech. {\bf 12}
(1997) 825.}
\bibitem{34} {A. Matulis, J. O. Fjarestad, and K. A. Chao, Physica Scripta 
{\bf T69} (1997) 85; {\bf T69} (1997) 138.}
\bibitem{35} {N. Akman and M. Tomak, Physica B, to be published.}
\bibitem{36} {K. Jauregui, W. Hausler, and B. Kramer, Europhys. Lett.
{\bf 24} (1993) 581.}
\bibitem{37} {X. -G. Li, W. -Y. Ruan, C. -G. Bao, and Y. -Y. Liu,
Few-Body Systems {\bf 22} (1997) 91.}
\bibitem{38} {J. H. Jefferson and W. Hausler, preprint (cond-mat) 9705012.}
\bibitem{39} {A. Brataas, U. Hanke, and K. A. Chao, Phys. Rev. {\bf B54} 
(1996) 10736.}  
\end{thebibliography}
\end{document}